\documentclass[journal,10pt,draftclsnofoot,onecolumn,compsoc]{IEEEtran}

\usepackage{mathtools}
\usepackage{graphicx}
\usepackage{tabulary}
\usepackage{multirow}
\usepackage[center]{caption}
\usepackage[hyphens]{url}

\ifCLASSOPTIONcompsoc
\else
\fi

\ifCLASSINFOpdf
\else
\fi

\begin{document}

\title{Trust-in-the-Middle: Towards Establishing Trustworthiness of Authentication Proxies using Trusted Computing }

\author{Yusuf~Uzunay,
		Kemal~Bicakci~
\IEEEcompsocitemizethanks{\IEEEcompsocthanksitem Y. Uzunay is with Middle East Technical University, Ankara, Turkey
E-mail: yusuf.uzunay@ua.gov.tr.
\IEEEcompsocthanksitem K. Bicakci is with TOBB University of Economics and Technology, Ankara, Turkey
E-mail: bicakci@etu.edu.tr.}
\thanks{}}

%

\IEEEcompsoctitleabstractindextext{%
\begin{abstract}
Authentication proxies, which store users' secret credentials and submit them to servers on their behalf, offer benefits with respect to security of the authentication and usability of credential management. However, as being a service that is not in control of users, one important problem they suffer is the trust problem; how users trust that their secret credentials are handled securely in the proxy and not revealed to third parties. In this paper, we present a solution called Trust-in-the-Middle, a TPM based proxy system which ensures that user credentials are securely stored and submitted without disclosing them even if the proxy is compromised. We build our architecture on a trust chain bootstrapped by TPM DRTM and we prevent access to credentials if any entity in the chain is maliciously modified. We use remote attestation to guarantee that all critical operations on the proxy are performed securely and credentials are cryptographically protected when they are not in DRTM-supported isolation.
\end{abstract}

\begin{keywords}
Authentication, Network-level security and protection.
\end{keywords}}

\maketitle

\IEEEdisplaynotcompsoctitleabstractindextext

%
\IEEEpeerreviewmaketitle

\section{Introduction}
%
%

%
%
%
%
\IEEEPARstart{E}{specially} in corporate networks, proxy systems are in use for variety of reasons such as caching, access control, content filtering, logging, etc. An application area for proxies that is not as popular as others but has received significant attention in the academic literature is to use them as agents for user authentication. With an authentication proxy, user first establishes a secure session with the proxy. Then, in each time user wants to login to a server, the proxy intercepts the connection, inserts the user credentials into the page and then submits it to the target server.

Two prominent advantages that authentication proxies can provide, improving the usability of credential management and increasing the security of user authentication, are described briefly as follows:

(i) Remembering and using large and continuously growing number of credentials (e.g., passwords, PINs and even usernames) becomes a real burden for users. Due to usability problems, users may prefer insecure options such as reusing the same password for different web sites. In this sense, authentication proxies enable users to store their credentials on the proxy and use them by entering  just a single password shared with the proxy.

(ii) Authentication proxies can also improve security by making it possible to use more secure alternatives such as one-time passwords even when the server itself does not support it.

In the literature, there is a considerable amount of work on authentication proxy systems (e.g., \cite{1,2,3,4,5,6}).  Although these proxy systems offer benefits with respect to security and usability, two of their problems are noteworthy; Firstly, authenticating users to the proxy system in a secure and usable fashion is still a serious problem. One may argue that the right balance for using more secure but less usable solutions like one-time passwords could be achieved by limiting their use only once per session opened with the proxy and only when an untrusted machine other than the user's primary computer is to be used. A one-time password based solution proposed in the earlier work on authentication proxy systems is also adopted in our proposed framework but we note that our contribution in this paper is not on this first problem.

The second problem, which is the central focus in our work, is less spoken but at least as serious as the first one; increasing the trustworthiness of the proxy system so that users would accept to hand over their sensitive credentials such as e-banking passwords to proxies without worrying about possible security breaches, intended or by mistake. This may be the main reason why proxy systems have not found a wide adoption among users for authentication purposes\footnote{A recent usability study confirms that users are not comfortable with giving control of their passwords to an online entity \cite{7}.}. In our literature survey, we see that previous work on proxy systems have made trust assumptions and this problem has not been studied in detail before.

In this paper, we make a first attempt towards establishing the trustworthiness of authentication proxies. For this purpose, we propose Trust-in-the-Middle, a proxy system based on trusted computing technology and its core component TPM (Trusted Platform Module). With the TPM Dynamic Root of Trust for Measurement (DRTM) functionality, we securely store and submit  sensitive credentials to the target servers without disclosing them even if the proxy server is compromised. All the security critical operations are carried out in the modules of which the integrity is protected by TPM. The credentials are never put out of DRTM protection without the cryptographic protection. Therefore, any malicious entity cannot intervene the operation and access the credentials in plaintext. We use remote attestation to verify the security of the software modules on the proxy. Sensitive data is sent to proxy only after the attestation result is checked.

The rest of the paper is organized as follows. Section 2 reviews the previous work.  Section 3 provides background information on TPM and its core functionalities used in the proposed work. We describe our proposed system, Trust-in-the-Middle, in detail in Section 4. Section 5 is for security analysis of the Trust-in-the-Middle system. We acknowledge that our work is only a first step towards a complete solution and thus is far from being perfect. The limitations in the current solution and our future work plans are presented in section 6 together with our concluding remarks.


\section{RELATED WORK}

Previous work is discussed under two headings: proxy-based systems and TPM-based systems.

\subsection{Proxy-based Systems}

A proxy-based system called Impostor for use from untrusted devices was proposed by Pashalidis and Mitchell \cite{1}. Impostor, the proxy, keeps a copy of user credentials for different web sites in this system. Whenever a user wants to connect to a site, Impostor intercepts the connection and sends a special login screen to the user. The login screen involves a challenge/response mechanism which requires users to share passphrases (at least eight characters) with the proxy server. The challenge asks user to provide three randomly chosen characters from the passphrase. If the response is correct, then Impostor sends the user's credential to the site and completes the authentication. By this way, if the user's machine is compromised, only a small portion of the secret i.e., the passphrase is revealed. As the challenge changes each time the user connects to the proxy, a replay attack is rendered to be more difficult. However, since the secret used for responses is same, an adversary obtaining an adequate number of responses is able to build the entire secret. The security of the proxy system is also not discussed for Impostor and an inherent trust assumption is made.	

Wu et al. \cite{2} propose another similar architecture where a proxy stores credentials and asks the user to respond to a challenge before submitting them. The challenge is also sent as an SMS message to the user's mobile phone. The SMS message includes a link which directs user to a WAP page to let him accept or deny the connection. By comparing challenges on two pages, user could avoid phishing attacks. In this system, the proxy is again assumed to be trusted.	

Delegate \cite{3} is another proxy based authentication system. As the main difference to the other systems we discuss, Delegate implements rule-based policies and requests additional credentials via the mobile phone of the user whenever a sensitive operation is to be carried out.

KLASSP \cite{4} proposed by Florencio and Herley is a proxy-based system which differs from other similar systems by not storing user passwords in the proxy. Although passwords are not stored on the proxy system, this does not eliminate the proxy trust problem because the proxy now holds other secrets (i.e., mapping table) to recover the password. Besides, a malicious software on the proxy system can access to the plaintext password while it is being submitted to the target server.

Florencio and Herley proposed another system based on the proxy idea \cite{5}. This time, one time passwords are also incorporated into the proposed solution. Before using the system, users register to the proxy called URRSA and provide the credentials (passwords) of the target servers.  URRSA generates the one-time passwords from the passwords using an encryption algorithm.	

Martineau and Kodeswaran proposed a very similar system to URRSA called SecurePass \cite{6}. However, the system has the same drawback with respect to trustworthiness of the proxy.

\subsection{TPM-based Systems}

Up to our best knowledge, there is no earlier work on applying trusted computing technology to authentication proxy systems. Below, we review previous work on using this technology for a more general problem, the problem of protecting credentials on untrusted environments.

One previous work that addresses the problem of protecting sensitive input on untrusted environments using TPM is Bumpy, proposed by McCune et al. \cite{8}. The system is based mainly on Flicker \cite{9}.  Bumpy allows the user to specify strings of input as sensitive while entering them and ensures that these input reach the desired endpoint in a protected state. The sensitive input are processed in an isolated code module (Flicker) on the user's system where they are encrypted or otherwise processed for a remote web server. Bumpy requires special equipment like an encrypting keyboard and may require change on the server side.

Li et al. proposed a secure user interface for web applications running on an untrusted operating system \cite{10}. With a small portion of code included in the user interface  a secure path from the user directly to the remote server is built. After the interface attests itself to the remote server, sensitive inputs are handled in this interface and transferred back to the OS with cryptographic protection. The proposed system utilizes TPM DRTM and Intel TXT technology to create an isolated environment as Flicker does \cite{9}. The difference is that a simple VGA and a keyboard driver are also added in the measured launch environment (MLE) (the isolated environment of Intel TXT technology created by TPM DRTM operation). By this way, users are able to interact with the MLE directly. When the user is required to input sensitive information, the browser places the MLE in the memory and invokes it to handle secure input and output.

Borders and Parkash proposed a virtualization based architecture \cite{11}. On the client machine, the keyboard inputs are directed to a trusted input-proxy (TIP) which executes in another virtual machine on the same machine. TIP replaces the real inputs with placeholders and sends them back to the primary OS. When the primary OS sends the packet to the network, the TIP searches for these placeholders and replaces them with the real inputs. T\_PIM\ \cite{12} is another similar solution which uses an activation password instead of placeholders and also uses TPM to verify the integrity of the trusted virtual machine. Both TIP and T\_PIM\ systems are deployed solely on client machines hence they are not proxy-based systems as the ones discussed in this paper. We also note that using virtual machines may bring additional security problems \cite{13}.

\section{BACKGROUND}
In this section, we provide background information on TPM and its core functionalities. Section 4 provides additional information on more advanced techniques regarding TPM use.

\subsection{Trusted Platform Module}
Trusted Platform Module (TPM), the core component of Trusted Computing \cite{14}, is a chip attached directly to the motherboard of the computer and stores keys, passwords and digital certificates. It has cryptographic capabilities such as RSA key generation, encryption, signing and verification, secure random number generation and SHA1 hashing.

The architecture of TPM includes a cryptographic processor, volatile and nonvolatile memories and secure input/output. In nonvolatile memory, endorsement key (EK), an RSA key pair used to uniquely identify the TPM, is located. This endorsement key is created by the manufacturer of the TPM at the time of production and the private key part of the EK never moves out of the TPM. TPM has also special registers called PCR (Platform Configuration Registers) which are used to store 160 bit SHA1 hash values. PCRs cannot be directly written. Instead, they are extended. TPM Extend is a special operation which calculates the new value of the PCR by hashing the concatenation of the old value and a new SHA1 hash value.

\subsection{TPM Trust Chain}
One of the important functions of TPM is forming a trust chain for which a set of entities are hashed and chained to each other by using TPM Extend operations. In practice, this functionality is often used to create a trust chain for software programs by verifying the integrity of all entities that have a potential to affect the trustworthiness of the software. This trust chain is also known as trusted computing base (TCB) of the software. For establishing the trust chain, two different methodologies have been developed; one of them is called SRTM (Static Root of Trust for Measurement) and the other is DRTM (Dynamic Root of Trust for Measurement).

In SRTM, TCB includes all the entities from system boot to the final software to be executed. The boot, also named as Trusted Boot, includes BIOS, Bootloader, operating system (OS) and all the drivers and software loaded till the execution of the final software. With SRTM, due to large TCB size, it becomes difficult to perform integrity measurements of all entities and form the trust chain reliably.

With TPM version 1.2, a new concept called DRTM has been introduced. DRTM avoids the disadvantages of SRTM and removes BIOS, bootloader, OS and other entities from the trust chain (a fresh boot is no more needed). With DRTM, CPU can reset the relevant PCRs at any time by using a specific instruction (SKINIT for AMD, SENTER for Intel) that atomically initializes the CPU, disables the interrupts and loads a piece of code into its cache. This code is sent to the TPM to be an input for TPM Extend operation and written on specific PCRs and then executed.  DRTM makes it possible to run a piece of code in an isolated environment which is not affected by any other entity on the computer system and stores the integrity measurements of the entities used during the DRTM operation on specific PCRs in order to provide the proof whether the relevant piece of code and all its components have been executed correctly.

For DRTM, AMD offers SVM-Secure Virtual Machine Technology \cite{15} and Intel offers TXT-Trusted eXecution Technology formerly LaGrande Technology (LT) \cite{16}.  (More information on TXT technology is presented in section 4).

\subsection{TPM Attestation Operation}

Attestation stands for proving the trustworthy status of a machine to a third party, which means that the machine has an original and enabled TPM and the requested hash values are correctly retrieved from the PCRs of the TPM chip. Basically, an attestation request includes a nonce and some PCR numbers. The attested machine then performs a TPM quote operation and produces a quote as a reply. This quote includes the signed values of nonce and the contents of the requested PCRs. The attested machine also sends an untrusted event log including the hash values of each entity that forms the trust chain in the relevant PCRs. The attester can then verify the untrusted event log by computing the aggregate hashes expected to be in the PCRs and compare the final value with the one in the quote signed by TPM \cite{9}.

Sign operation is performed via the private portion of a special key, called Attestation Identity key (AIK). The security of AIK key is bootstrapped from the TPM's EK (Endorsement key) which is unique for each TPM. In the attestation process, Privacy CA is a trusted third party certifying that the AIK is generated by a legitimate TPM. There is another attestation type called direct anonymous attestation (DAA) \cite{17} which enables trusted computers to attest directly and anonymously without using a third party. However due the complexity of DAA, most work prefers using a Privacy CA.

\subsection{TPM Seal Operation}
TPM Seal operation, another important function provided by TPM, bounds the encrypted message with a non-migratable private key of TPM and contents of selected PCR values. By this way, it is guaranteed that the encrypted message can only be decrypted by the TPM which performed the encryption and when the contents of the relevant PCRs are as same as the contents during the encryption operation.

\section{PROPOSED SYSTEM}
Our proposed system Trust-in-the-Middle is explained in the following sections.

\subsection{Model, Objectives and Assumptions}

\begin{figure}[!htb]
\centering
\includegraphics[width=2.5in]{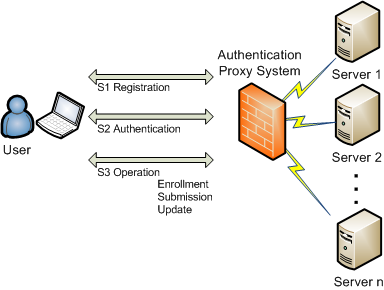}
\caption{System Model of our Authentication Proxy System}
\label{fgr:Fig1}
\end{figure}

Before giving the technical details of Trust-in-the-Middle system, we present our system model. We also describe the assumed attacker model informally and explain our objectives, limitations and assumptions:

{\bf System Model.} Figure~\ref{fgr:Fig1} depicts the system model of our authentication proxy system. There are three entities: (i) user, (ii) proxy, (iii) target servers. The user, who wants to authenticate himself to several target servers, uses the authentication proxy system which is in a relay agent position intercepting the communication and carrying out the tasks required for authentication on behalf of him.

Within this system model, there are three main services provided by the proxy: (S1) Registration, (S2) Authentication and (S3) Operation.

{\it S1 Registration.} The user registers to the system and shares with the proxy a master password and optionally a one-time password list.

{\it S2 Authentication.} User's identity is verified first by the proxy before serving him for further operations.

{\it S3 Operation.} The following operations are made available by the authentication proxy system:

\begin{itemize}
\item Enrollment: Users could enroll their credentials for the servers on the authentication proxy at any time.
\item Submission: Credentials, stored on the proxy, could be submitted to the servers on behalf of their authenticated owners.
\item Update: Users could also update their stored credentials on the proxy any time they wish.
\end{itemize}

{\bf Attacker Model.} We assume that the main goal of an attacker is to obtain user credentials. We can list four places where credentials are under threat: (i) Client machine, (ii) Network between clients and the proxy as well as between the proxy and target servers, (iii) Servers, (iv) Authentication proxy.

{\it Client machine.} The attacker may try to capture user credentials while they are being entered on the client machine by using methods such as keystroke logging, screen-scraping, malicious code injection on the operating system, browser or any other software on the client.

{\it Network between client-proxy-servers.} The attacker may conduct several types of attacks in order to capture the credentials on the network i.e. network sniffing, pharming, man-in-the-middle attacks, etc.

{\it Target Servers.} The attacker can perform an attack on server machines to capture the stored credentials.

{\it Authentication Proxy.} The attacker may try the following methods to obtain credentials while they are being processed or stored on the proxy: He can attack directly to the database where the credentials are stored and try password guessing or dictionary attacks if the credentials are stored in hashed form. He can try to capture the credentials while they are being enrolled, submitted or updated. He can inject malicious code on the software modules running the enrollment, submission and update protocols.

{\bf Security Objectives and Limitations.} Our main security objective in this paper is to focus on authentication proxy systems and address the concerns with respect to security of user credentials while they are being stored and processed on these systems. We take into consideration each services provided by the proxy and provide an overall security architecture.

Despite being not our focus, we note that our proposed framework also provides protection against some of the client side attacks (due to its support for one-time passwords) and network attacks (due to use of encrypted tunneling). See the discussion in Section 5 for more details.

A limitation of the proposed work is that we are concerned with the security of the authentication prior to proceeding with the online transaction, not with the security of the transaction as a whole. Hence sophisticated attacks such as session hijacking attacks, transaction generators, etc. are not addressed. Furthermore, attacks to target servers, denial of service attacks, physical attacks, social engineering attacks are also out of scope in our work. We also do not discuss the privacy implications of using an authentication proxy.

{\bf Usability Objectives and Limitations.}	Our objectives are listed as follows:

\begin{itemize}
\item To enable users to access different password protected web sites using only one (master) password. As a result, users do not need to memorize multitudes of passwords.
\item To provide users an easy access to the password protected web sites as long as their authentication session with the proxy system is alive. After authenticated to the proxy, users could feel as if they were not using passwords at all
\item To provide a smooth user experience not only during login but also for password update i.e., by filling the current password fields automatically while user is changing password and by updating stored passwords on the proxy at the background without requiring further user action.
\item The proposed system should be transparent; not requiring any change on server side
\item The proposed system should not require a specific architecture or operating system on the user side.
\end{itemize}

The usability limitations of the proposed system are as follows:
\begin{itemize}
\item There is a need for registration to the proxy before using the system (but the registration does not need to be offline).
\item Users should install a special client software on their machines to use the authentication proxy service.
\item Users should make relevant proxy settings in their browsers\footnote{We present an add-on in our prototype implementation that makes the proxy settings automatically for using the Trust-in-the-Middle system.}.
\end{itemize}

{\bf Additional Assumptions.} We also make the following assumptions:
\begin{itemize}
\item Proxy machine has a TPM v1.2 chip and support TPM DRTM.
\item Source codes, binary files and hash values of software modules used in the proxy system are publicly available and verified as being secure.
\item The client software is not compromised.
\end{itemize}

\subsection{Overview}
In this section, we present an overview of how our proposed system, Trust-in-the-Middle, implements the system model given in Figure \ref{fgr:Fig1}.

Before using Trust-in-the-Middle, a user needs to register to the system and obtain a proxy authentication master password and optionally a one-time password list. This is carried out through a client software on user side which runs a secure registration protocol. Registration requires to be performed on trusted client machines.

After completing the registration and after configuring the relevant proxy settings, the user can start using the system. Through the client software first he authenticates himself to Trust-in-the-Middle. Here, user has two options; using a master password or a one-time password. If user trusts his platform (for instance if it is his primary machine), he can use the master password. Otherwise, one-time passwords may be preferred. During authentication, a remote attestation protocol is executed between client software and the proxy system and only if the connection and the connected module is verified to be secure, user sends his password to the proxy. If the authentication is successful, a tunnel is established between the client and the proxy system. This tunnel is used to open a local port on client system and redirect the coming traffic to the Trust-in-the-Middle proxy service.


User can now visit any web site he wants to login. Since proxy settings of the user are configured to forward all the web traffic to the Trust-in-the-Middle through the established tunnel, Trust-in-the-Middle intercepts the connection. As most of the security critical login pages require SSL, Trust-in-the-Middle functions as an SSL MITM proxy and sets up two independent SSL connections, one with the user and one with the target server.  While establishing the SSL connection with the server, Trust-in-the-Middle checks its SSL certificate and establishes connection only if it can verify the certificate correctly. Trust-in-the-Middle then checks whether the authenticated user has previously stored credentials for the target server or not. If it finds a match, it sends the login page to the user by filling in the credential fields with dummy credentials. Otherwise, it sends the page with empty credential fields.

Seeing that the login page is prefilled, user understands that Trust-in-the-Middle has filled it for himself. So the only thing he needs to do is to click on the submit button. Receiving this submission, Trust-in-the-Middle first retrieves the encrypted credentials from user database and initiates a TPM DRTM operation to securely decrypt and obtain the credentials. After that, Trust-in-the-Middle inserts the real user credentials into the correct fields and submits the page to the target server.

If the login page has empty credential fields, user understands that he has not entered these credentials using Trust-in-the-Middle before. If the user trusts the client environment, he enters the credentials so that Trust-in-the-Middle submits them on behalf of him and also enrolls them on proxy for future use.

Trust-in-the-Middle also supports credential update operation. When the user visits the credential update page of the target web site, Trust-in-the-Middle is able to detect and fill in the current credential fields automatically. For the new credentials, the same process as credential enrollment is followed. Trust-in-the-Middle receives the new credentials, encrypts them with TPM protected keys and then updates its user database.


\begin{table*}[!hbt]
{
\begin{center}
\caption{Protocols used for implementing the services of the Trust-in-the-Middle system}
\label{TITM_Protocols}
\centering
\renewcommand{\arraystretch}{1.4}
\begin{tabular}{|c|c|c|c|}
\hline
\multirow{2}{*}{\textbf{Services}} & \multirow{2}{*}{\textbf{Main Protocols}} & \multicolumn{2}{c|}{\textbf{Auxiliary Protocols}} \\
\cline{3-4}
& & \textbf{Level-1} & \textbf{Level-2} \\
\hline
Registration & Registration Protocol & Secure Tunnel Protocol & Attestation Protocol\\
\hline
Authentication & Authentication Protocol & Secure Tunnel Protocol & Attestation Protocol\\
\hline
Operation(Enrollment) & Credential Enrollment Protocol & Credential Decryption Protocol & - \\
\hline
Operation(Submission) & Credential Submission Protocol & Credential Decryption Protocol & - \\
\hline
Operation (Update) & Credential Update Protocol & Credential Decryption Protocol & - \\
\hline
- & Initial Sealing Protocol & - & - \\
\hline
\end{tabular}
\end{center}
}
\end{table*}


Trust-in-the Middle provides the described services by executing  Main and Auxiliary protocols given in Table \ref{TITM_Protocols}.  Main protocols execute the auxiliary protocols at level 1 which may execute another auxiliary protocol at level 2.

In Table \ref{TITM_Protocols}, there is a specific main protocol, Initial Sealing Protocol, which does not have a match with a service or an auxiliary protocol. This protocol is executed only once when the Trust-in-the-Middle system boots. It is used to protect the integrity of the public key of the proxy module with the help of TPM till the next boot of the system. Other main protocols are responsible for carrying out  the operations regarding five different services; Registration protocol is used to create and store a master password and a one-time password list to be used in subsequent proxy authentication. Authentication protocol is used to authenticate users with previously registered master or one-time passwords. Credential enrollment protocol is responsible for enrolling user credentials in Trust-in-the-Middle database encrypted with TPM protected keys. Credential submission protocol is used to decrypt and insert user credentials into the login page and perform submit operation. Credential update protocol is used to replace previously stored credentials by the new ones.

Main protocols use two auxiliary protocols at level 1, Secure Tunnel Protocol and Credential Decryption Protocol. The main job of secure tunnel protocol is to establish a tunnel with the security sensitive code running in TPM protected environment. By this way a direct and secure communication with the sensitive code is provided. The Credential Decryption Protocol is used to obtain credentials securely which were previously encrypted with TPM protected keys. The only auxiliary protocol running at level 2 is attestation protocol which is executed by secure tunnel protocol in order to verify whether the correct code is executed in TPM protection.

\begin{figure*}[!hbt]
\begin{center}
\centering
\includegraphics[width=5.0in]{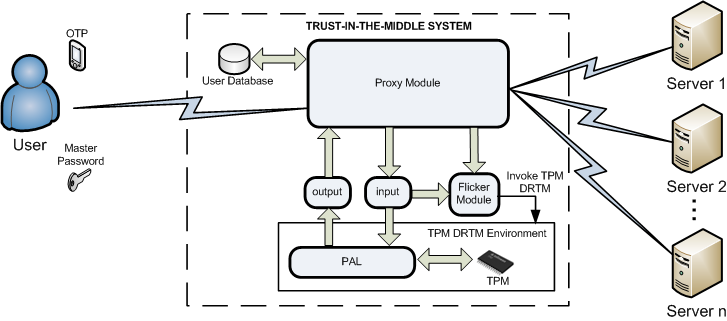}
\caption{Trust-in-the-Middle System Architecture}
\label{fgr:Fig2}
\end{center}
\end{figure*}

In the following sections, we first give an overall architecture of the system (Figure \ref{fgr:Fig2}) and explain the role of each entity in this architecture. Then, we introduce the code structure of security sensitive code named as PAL-Program Application Logic to be executed in TPM and explain each function of the code. Finally, we give the details of each protocol used in the proposed system and explain their operation.

\subsection{Architecture and Technology}
In this section, we explain the architecture of Trust-in-the-Middle system together with the technologies used. Abbreviations used in the architecture and the protocols are presented in Table \ref{Abbreviations1} and Table \ref{Abbreviations2}, respectively.

\begin{table}[!htb]
\caption{List of Architecture Abbreviations}
\label{Abbreviations1}
\centering
\renewcommand{\arraystretch}{1.2}
\begin{tabular}{|l|l|}
\hline
AC & Authenticated Module\\
\hline
AIK & Attestation Identity Key\\
\hline
CA & Certificate Authority\\
\hline
DRTM & Dynamic Root of Trust for Measurement\\
\hline
MITM & Man-in-the-Middle\\
\hline
MLE & Measured Launch Environment\\
\hline
PAL & Program Application Logic\\
\hline
PCR & Platform Configuration Register\\
\hline
SKINIT & Secure Kernel Init\\
\hline
SLB & Secure Loader Block\\
\hline
SML & Stored Measurement Log\\
\hline
SRK & Storage Root Key\\
\hline
TCB & Trusted Computing Base\\
\hline
TPM & Trusted Platform Module\\
\hline
TXT & Trusted Execution Technology\\
\hline
PM & Proxy Module\\
\hline
DB & Database\\
\hline

\end{tabular}
\end{table}

\newcommand{\specialcell}[2][c]{%
  \begin{tabular}[#1]{@{}l@{}}#2\end{tabular}}

\begin{table}[!htb]
\caption{List of Protocol Abbreviations}
\label{Abbreviations2}
\centering
\renewcommand{\arraystretch}{1.1}
\begin{tabular}{|l|l|}
\hline
PALpriv & Private Key of PAL\\
\hline
PALpub & Public Key of PAL\\
\hline
PMpriv & Private Key of Proxy Module\\
\hline
PMpub & Public Key of Proxy Module\\
\hline
MasPass & Master Password\\
\hline
SecPhrase & Secret Phrase used to generate one time passwords\\
\hline
PassList & List that holds proxy authentication passwords\\
\hline
sealedPassList & Sealed Password List\\
\hline
DumCred & Dummy Credentials\\
\hline
NewCred & New credentials in update\\
\hline
OldCred & Old credentials in update \\
\hline
sealedPALPriv & Sealed Private Key of PAL\\
\hline
sealedPMPub & Sealed Public Key of Proxy Module\\
\hline
encCredwithPal & Encrypted Credentials with PAL public key\\
\hline
encCredwithPM & Encrypted Credentials with PM public key\\
\hline
encNewCredwithPal & Encrypted New Credentials with PAL public key\\
\hline
SenData & Sensitive Data\\
\hline
EncSenData & Encrypted sensitive data with PAL public key\\
\hline
OTP & One-Time Password\\
\hline
H() & Hash Operation\\
\hline
Enc() & Encryption Operation\\
\hline
Dec() & Decryption Operation\\
\hline
Cert(AIKpub) & Certificate of AIK public key\\
\hline
Sig() & Signing Operation\\
\hline
\end{tabular}
\end{table}

The architecture of Trust-in-the-Middle is illustrated in Figure \ref{fgr:Fig2}. There are six main components in this architecture; Proxy Module, PAL (Program Application Logic), Flicker Module, User Database, Input File and Output File.

Proxy module takes an important role in all system services. It supports SSL MITM functionality and incorporates the engines implemented to detect the login and update services of web sites and to insert user credentials into the correct fields. An important functionality of proxy module is to manage the communication with PAL by invoking TPM DRTM environment with the help of Flicker module. PAL is a portion of code responsible for performing security critical operations in TPM protection. Whenever a TPM DRTM environment is required for executing the PAL, proxy module invokes Flicker module, responsible for preparing the relevant environment and the operating system to run TPM DRTM. PAL is executed in an isolated environment provided by TPM DRTM. It is not possible to communicate directly with PAL through user level modules during its execution. The only way to communicate with PAL at this phase is using input and output files.  User database holds the information about users and servers, encrypted credentials of users and proxy authentication passwords.

Two important technologies used in the architecture of Trust-in-the-Middle are described next.

{\bf TPM DRTM with Intel TXT Technology:} As the platform supporting TPM DRTM operations, we use Intel Trusted Execution Technology (TXT). The platform establishes the authenticity of a measured launched environment (MLE) and protects this environment from potential corruption. MLE then establishes an isolated environment for itself and additional software it may execute \cite{16}.

In order to be able to launch MLE, first of all, an Authenticated Module (AC), which is specific for the chipset and digitally signed by the chipset vendor, should be loaded into the memory. Only when AC module and MLE are in memory, the launching environment can invoke an instruction (i.e., GETSEC[SENTER]) which initiates the TPM DRTM functionality on the processor. This specific command brings the chipset and CPU in a stable state and loads, validates and executes the AC. AC module then ensures that the platform has a proper configuration and measures and launches the MLE \cite{10}. These measurements are stored in specific PCRs using TPM Extend operations.  In our proposed system, the security critical code (Program Application Logic - PAL) runs on this launched environment.

{\bf Flicker:} Intel TXT offers capabilities to use TPM DRTM environment and run security sensitive code on it. As we mentioned, TPM DRTM is only invoked when the relevant AC Module and the MLE are in the memory. In addition, before DRTM invocation there are other requirements such as preparing the PAL and locating it on the right section in the memory and backing up the system state in order to resume it after PAL has finished its job and creating a directory structure to provide data exchange with PAL. For all these purposes, we use Flicker developed by McCune et.al. \cite{9}.

Flicker is written as a SYSFS module which is a virtual file system capable of exporting information about devices and drivers from kernel to user space so that it becomes possible to make data exchange between a user level application and the Flicker module.

Whenever a DRTM operation is needed, Flicker module is loaded into the kernel and invoked with input parameters and SLB (Secure Loader Block \cite{15}) which includes the PAL (Program Application Logic). Then, Flicker suspends the OS by saving the existing state and gives control to the loaded SLB by executing a DRTM command\footnote{Flicker version 0.1 is based on AMD SVM technology. We use Flicker 0.2 version which supports Intel TXT hence TPM DRTM command in our solution is GETSEC[SENTER].}. After SLB has completed its work and the relevant PCRs are extended, Flicker takes control again and resumes the operating system. PALs can use TPM-based sealed storage to maintain state across Flicker sessions, enabling more complex applications \cite{9}.

\subsection{PAL Overview}
PAL (Program Application Logic) is a name we adopt from Flicker \cite{9} and use it to refer to the security sensitive code executed in TPM DRTM protection.	

Since TPM DRTM operation provides a restricted environment where all the interrupts are disabled, the complex software programs requiring user-level operations such as Proxy Module in our application are not included directly in PAL.  We prefer keeping the PAL as small as possible by including the most critical parts of the operations (i.e., Integrity Measurements, Seal/Unseal operations and Extend operations) and leave the other user level operations to other modules.

PAL is executed by Flicker module \cite{9} invoked from user space. After Flicker prepares the required environment, it invokes TPM DRTM and gives the control to PAL. PAL execution is carried out in a secure and isolated environment which cannot be accessed from user space. For this reason a user level application cannot communicate directly with PAL during the TPM DRTM session. The input data for PAL has to be written in a specific input file of Flicker which is then located in a known address in the memory. If PAL has any output after its execution, this is also written in a known address in the memory and made available to user level applications via Flicker output file.  Since these input and output files can be accessed by any user level application, the data is encrypted before written down on these files.

\begin{figure}[!htb]
\centering
\includegraphics[width=2.5in]{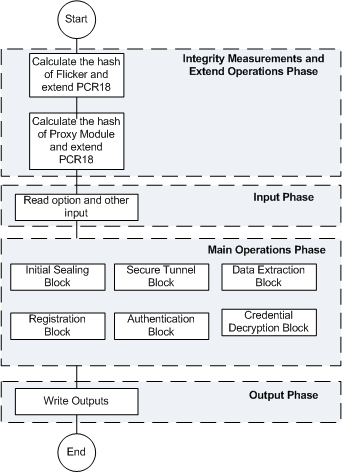}
\caption{PAL Overview}
\label{fgr:Fig3}
\end{figure}

An overview of PAL implemented in our system is presented in Figure \ref{fgr:Fig3}. The operation of PAL is performed in four phases: Integrity Measurements and Extend Operations Phase, Input Phase, Main Operations Phase and Output Phase.

{\bf Integrity Measurements and Extend Operations Phase.} The security of the Trust-in-the-Middle system depends on the relationship between PAL, Proxy and Flicker modules. PAL is the main software that is responsible for checking the integrity of other software modules and making the seal and unseal operations. PAL is executed in TPM DRTM protection and the hash of PAL is extended into PCR18. Therefore, the integrity measurement of PAL is directly provided by TPM. The integrity measurements and Extend operations of the other modules are performed by the PAL (see Figure \ref{fgr:Fig3}). As a result, other modules are also added into the TCB (Trusted Computing Base). By verifying the integrity of the TCB, user credentials on Trust-in-the-Middle is protected against malicious infection.

To establish the integrity of the modules, we set up a trust chain utilizing TPM Extend operations. Furthermore, we use this chain and TPM Seal operation to protect the private portion of the key pair used in encryption and decryption operations for user credentials. The SML (stored measured log) of our trust chain is given below:

SML $\leftarrow$ \{PAL, Flicker Module, Proxy Module\}	

Whenever a TPM Seal/Unseal operation is needed, TPM DRTM  environment is invoked with the PAL. After preparing the environment and loading the MLE securely, TPM sets the value of PCR18 to "0", extends it with the hash of PAL and then starts executing the PAL code. All operations till this point are the standard operations of TPM DRTM functionality. PAL then performs two more Extend operations for Flicker module and Proxy Module (PM). The final value of PCR18 is determined by the following hash chain:	

PCR18 $\leftarrow$ H(H(PM)+H(H(Flicker)+H(H(PAL)+"0")))

Using this trust chain and TPM Seal and Unseal operations, we ensure that the extraction of sealed data (i.e., sealed private keys) succeeds only if the integrity measurements of PAL, Flicker and PM are verified to be correct.

\begin{table*}[!htb]
{
\begin{center}
\caption{Executed PAL Blocks in Main Operation Phase corresponding to Trust-in-the-Middle services}
\label{ExecutedPALBlocks}
\centering
\renewcommand{\arraystretch}{1.4}
\begin{tabular}{|c|c|c|c|}
\hline
\textbf{Service} & \multicolumn{3}{c|}{\textbf{Executed PAL Blocks}} \\
\hline
Registration & Registration Block & Secure Tunnel Block & Data Extraction Block\\
\hline
Authentication & Authentication Block & Secure Tunnel Block & Data Extraction Block\\
\hline
Operation(Enrollment) & Credential Decryption Block & - & - \\
\hline
Operation(Submission) & Credential Decryption Block & - & - \\
\hline
Operation (Update) & Credential Decryption Block & - & - \\
\hline
\end{tabular}
\end{center}
}
\end{table*}

{\bf Input Phase.} After Integrity Measurements and Extend Operations Phase, PAL reads the input (input file was retrieved and located to a specific memory location by Flicker Module). We note that there is a specific input called option indicating the operation block to be invoked by PAL.

{\bf Main Operations Phase.} There are six main operations that can be executed by PAL according to the input option value. Initial Sealing Block is used to seal the public key of Proxy Module during the trusted boot.  Secure Tunnel Block is used to establish a secure tunnel between PAL and a remote entity. Data Extraction Block is used to extract the data received through the secure tunnel. Registration Block is used to create and store proxy authentication passwords protected by TPM Seal operation.  Authentication Block is used to carry out the proxy authentication with the credentials received from user. Credential Decryption Block is used to unseal the private key used in credential encryption previously and decrypt the credentials with the unsealed private key.

{\bf Output Phase.} At this phase, if it is needed, PAL writes the output to a specific memory location which is then written into the output file.

Integrity Measurements and Extend Operations, Input and Output Phases are usually executed for every PAL invocation. However the blocks executed in Main Operation Phase differ according to the service type. Table \ref{ExecutedPALBlocks} shows executed PAL blocks corresponding to each service.

\subsection{Auxiliary Protocols}
In this section, we describe the operation of auxiliary protocols used by the main protocols of Trust-in-the-Middle system (see Table~\ref{TITM_Protocols}).

{\bf Attestation Protocol:} Attestation has a crucial role in the proposed system and is used to verify that the correct PAL is executed before establishing the secure tunnel. Attestation Protocol is given in Table \ref{Attestation_Protocol}.

\begin{table}[!htb]
\caption{Attestation Protocol}
\label{Attestation_Protocol}
\centering
\renewcommand{\arraystretch}{1.2}
\begin{tabular}{|l l|}
\hline
1a.Verifier		   &       : Generate nonce \\
2a.Verifier$\rightarrow$Attestor  &        : Attestation Request (nonce, PCRno)\\
3a.Attestor	      &    : Loadkey(AIKkey)\\
4a.Attestor$\rightarrow$TPM     &          : Execute TPM Quote Operation\\
5a.TPM &	                            : Quote=sig\{PCR,nonce\}AIKpriv\\
6a.TPM$\rightarrow$Attestor &              : Quote\\
7a.Attestor                         &  : Generate (SML)\\
8a.Attestor$\rightarrow$Verifier	 &         : Quote, SML and cert(AIKpub)\\
9a.Verifier	     &     : validate cert(AIKpub)\\
9b.Verifier   & : validate sig\{PCR,nonce\}AIKpriv\\
9c.Verifier	     &     : validate SML using PCR\\

\hline

\end{tabular}
\end{table}

Attestation Protocol starts with a nonce value generation. This nonce value has an important role in providing the freshness of the attestation and preventing replay attacks. Verifier (the client) then sends an attestation request to the attestor (the proxy) with the nonce and a PCR number indicating which PCR is to be used in attestation. Receiving this request, the attestor needs to perform a TPM Quote operation. For this purpose, an AIK Key is loaded into the TPM slot first. This AIK Key is an encrypted key bound with Storage Root Key which is a non-migratable key embedded in the TPM nonvolatile memory. Extraction of private key from AIK Key can only be done inside the TPM and cannot be accessed from outside. After loading AIK Key into the TPM, attestor performs TPM Quote operation and obtains a Quote which is formed by concatenated PCR and nonce values signed by AIK private key. Then, attestor creates the SML (the hashes of each entity creating the trust chain), and sends SML, Quote and the AIK certificate to the verifier. The verifier validates the AIK certificate, the signature and the nonce value. Then, it calculates the final hash value from the SML and validates the PCR value.

{\bf Secure Tunnel Protocol:} In order to send sensitive data directly to the PAL, the client establishes a secure tunnel using the protocol given in Table \ref{Secure_Tunnel_Protocol}.

\begin{table}[!htb]
\caption{Secure Tunnel Protocol}
\label{Secure_Tunnel_Protocol}
\centering
\renewcommand{\arraystretch}{1.2}
\begin{tabular}{|l l|}
\hline
1a.Client$\rightarrow$PM &	: Secure Tunnel Request\\
2a.PM$\rightarrow$PAL &	: Invoke PAL with "Secure Tunnel Block"\\
3a.PAL &		 : Generate an RSA keypair (PALpriv,PALpub) and nonce\\
3b.PAL &		: sealedPALpriv=TPMSeal(\{PALpriv, nonce\},PCR18)\\
3c.PAL &		: TPMExtend(PCR18,sha1(PALpub,nonce))\\
4a.PAL$\rightarrow$PM &	: PALpub, sealedPALpriv, nonce\\
5a.PM$\rightarrow$Client &   	: PALpub, nonce\\
6a.Client &   : Execute Attestation Protocol and validate PALpub and nonce\\
7a.Client &                   : encSenData=Enc(SenData)PALpub\\
8b.Client$\rightarrow$PM &  	: encSenData\\
9a.PM$\rightarrow$PAL &  	: Invoke PAL with "Data Extraction Block"\\
9b.PM$\rightarrow$PAL &       	: encSenData, sealedPALpriv, nonce\\
10a.PAL &         : \{PALpriv,nonce\} = TPMUnseal(sealedPALpriv,PCR18)\\	
10b.PAL &                     : Validate nonce values\\
10c.PAL &                     : SenData=Dec(encSenData)PALpriv\\
\hline
\end{tabular}
\end{table}

Receiving a secure tunnel request, PM invokes a PAL session and initiates a Secure Tunnel Block. PAL generates an RSA key pair and a nonce value. PAL then seals PAL private key and nonce with PCR18 and makes a TPM Extend operation to the PCR18 with the hash of PAL public key and nonce. PAL ends its session by providing PAL public key, sealed PAL private key and nonce values as output. 	 PM sends PAL public key and nonce value to the client. Client first executes Attestation Protocol to verify that the correct PAL has been executed and the output values of the PAL are correct. If the verification is successful, it encrypts the sensitive data (e.g., the master password) with PAL public key and sends it to the PM.

PM invokes another PAL session and initiates Data Extraction Block. PM also provides sealed PAL private key, encrypted sensitive data and nonce as input. Upon receiving these input, PAL performs a TPM Unseal operation to recover the PAL private key and the nonce. This operation succeeds only if the value of PCR18 is as same as the value in the previous PAL session ensuring the integrity of the PAL. After the unseal operation, if the nonce value is correct, PAL decrypts encrypted sensitive data with its PAL private key. 	

With this Secure Tunnel Protocol, we prevent any malicious entity between the client and the PAL to access the sensitive data in plaintext.

{\bf Credential Decryption Protocol:} Credential Decryption Protocol, given in Table \ref{Credential_Decryption_Protocol}, is an auxiliary protocol called by Credential Enrollment, Submission and Update Protocols to decrypt previously encrypted user credentials using a TPM protected (sealed) private key.


\begin{table}[!htb]
\caption{Credential Decryption Protocol}
\label{Credential_Decryption_Protocol}
\centering
\renewcommand{\arraystretch}{1.2}
\begin{tabular}{|l l|}
\hline
1a.PM &	                  : Generate \emph{nonce'}\\
2a.PM$\rightarrow$PAL &	: Invoke PAL with "Credential Decryption Block"\\
2b.PM$\rightarrow$PAL &	: encCredwithPAL, sealedPALpriv, nonce, sealedPMPub, \emph{nonce'}\\

3a.PAL &		: PMPub=TPMUnseal(sealedPMPub,PCR18)\\
3b.PAL &		: \{PALpriv, nonce\}=TPMUnseal(sealedPALpriv, PCR18)\\
3c.PAL &		: Validate nonce value\\
3d.PAL &		: Credentials=Dec(encCredwithPAL)PALpriv\\
3e.PAL & : encCredwithPM=Enc(Credentials,\emph{nonce'})PMPub\\
4a.PAL$\rightarrow$PM &	: encCredwithPM\\
5a.PM &		: \{Credentials, \emph{nonce'}\}=Dec(encCredwithPM)PMpriv\\
5b.PM &		: Validate \emph{nonce'} value\\
\hline

\end{tabular}
\end{table}

In the protocol, Proxy Module first generates a nonce value (\emph{nonce'}) and invokes a PAL session by initiating the Credential Decryption Block. It then sends previously encrypted credentials, sealed PAL private key and its nonce value, sealed public key of proxy module and the nonce value (\emph{nonce'}) generated at the beginning of the protocol to the PAL as input. Note that there are two different nonce values used in this protocol. The first nonce value (nonce) is the input to the PAL that was previously used in PAL private key sealing operation and is needed to validate the private key after unseal operation. The second value (\emph{nonce'}) is used in the encryption of data sent to Proxy Module with the goal of preventing replay attacks.

Receiving the input, PAL first unseals the sealed public key of Proxy Module (the public key of proxy is sealed by PAL during the trusted boot process discussed in Section 4.6). Then, PAL unseals the sealed PAL private key and obtains the private key and the nonce value. If the nonce is correct, it decrypts the encrypted credentials with PAL private key. After obtaining the credentials in plaintext, PAL first concatenates the credentials with the nonce' value and encrypts them with the unsealed public key of proxy module. PAL then sends them to Proxy Module. Receiving this, Proxy Module performs a decryption operation by using its private key and accesses the credentials and the \emph{nonce'} in plaintext. It validates \emph{nonce'} before using the plaintext credentials.

\subsection{Main Protocols}
In this section, we explain the main protocols used in registration, authentication, credential enrollment, update and submission services (see Table \ref{TITM_Protocols}). We describe the protocol used in initial sealing, first.

{\bf Trusted Boot and Initial Sealing:} In Trust-in-the-Middle system, Proxy Module and Flicker Module start running as a service when the system is booted. However, PAL is not a service running continuously. The integrity of the modules running as a service is provided by a trusted boot operation. With the help of Intel TXT and tboot \cite{18}, we implement a trusted boot and prevent booting when one of the hash values in the boot chain is changed.  Since we use DRTM, we do not need to include the hash values of other entities in our trust chain, it just contains the hashes of Proxy Module and the Flicker Module.


\begin{table}[!htb]
\caption{Initial Sealing Protocol}
\label{Initial_Sealing_Protocol}
\centering
\renewcommand{\arraystretch}{1.2}
\begin{tabular}{|l l|}
\hline
1a.PM &		: Generate an RSA keypair (PMpriv,PMpub)\\
1b.PM &		: TPMExtend(PMpub,PCR15)\\
1c.PM &		: Validate the hash of PAL\\
2a.PM$\rightarrow$PAL &	: Invoke PAL with "Initial Sealing Block"\\
2b.PM$\rightarrow$PAL &	: PMPub\\
3a.PAL &		: hash=Hash("0"+Hash(PMPub))\\
3b.PAL &		: validate PMPub by checking hash=PCR15\\
3c.PAL &		: sealedPMPub=TPMSeal(PMPub,PCR18)\\
4a.PAL$\rightarrow$PM &	: sealedPMPub\\
5a.PM &		: Store sealedPMPub\\
\hline

\end{tabular}
\end{table}

Proxy Module starts its operation by executing the initial sealing protocol given in Table \ref{Initial_Sealing_Protocol}. In this protocol, first an RSA key pair is generated and the hash of public portion is extended into one of the empty PCRs (PCR15 in our implementation) which has a default value "0". This PCR is used by PAL in order to verify the public key of the Proxy Module.

The hash of PAL is validated by Proxy Module and if it is correct, PAL is invoked for an initial sealing operation using the proxy public key. Receiving the public key, PAL first calculates the hash of public key, then concatenates it with "0" and again performs the final hash operation. If the calculated value matches with the PCR15 value, it ensures that public key belongs to the proxy module. It performs TPM Seal operation on the public key and the sealed public key is stored by the Proxy Module for later use.

The private key of the Proxy Module is not written to a file and kept in the memory as long the Proxy Module runs as a service (see section 5 for the security issues herein).

{\bf Proxy Registration:} If proxy authentication password is compromised, all credentials enrolled in the Trust-in-the-Middle becomes vulnerable. Therefore, we assume users perform registration only using secure platforms. The protocol in Table \ref{Registration_Protocol} is executed for proxy registration.


\begin{table}[!htb]
\caption{Registration Protocol}
\label{Registration_Protocol}
\centering
\renewcommand{\arraystretch}{1.2}
\begin{tabular}{|l l|}
\hline
1a.Client$\rightarrow$PM &	: Registration Request\\
2a.PM$\rightarrow$PAL &	: Invoke PAL with "Registration Block"\\
2b.PM$\rightarrow$PAL &	: sealedPassList\\			
3a.Client$\leftrightarrow$PAL &	: Execute Secure Tunnel Protocol\\
3b.Client$\rightarrow$PAL &	: Userid, MasPass and SecPhrase\\
4a.PAL &		: Generate OTP passwords (OTP[list])\\
4b.PAL &		: PassList= TPMUnseal(sealedPassList,PCR18)\\
4c.PAL &		: Update PassList with Userid, MasPass and OTP[list]\\
4d.PAL &		: sealedPassList=TPMSeal(PassList,PCR18)\\
5a.PAL$\rightarrow$PM &	: OTP Parameters, sealedPassList\\
6a.PM$\rightarrow$DB &	: sealedPassList\\
7a.PM$\rightarrow$Client &	: OTP Parameters\\
8a.Client &	: Generate OTP[list] by using OTP Parameters and SecPhrase\\
\hline

\end{tabular}
\end{table}

Upon receipt of a registration request, Proxy Module invokes a PAL session, initiates a Registration Block and gives sealed password list to the PAL as input.

The Secure Tunnel Protocol is executed between PAL and the client in order to establish a secure tunnel. User determines a master password and a secret phrase which will be used in the generation of one time passwords. As we mentioned before, Trust-in-the-Middle offers two password options for the user, master password and one-time passwords. PAL runs an OTP generation algorithm with the given secret phrase to generate a list of one-time passwords.

The sealed password list, including all user IDs and passwords of all registered users, is TPM protected.  Hence, for a new registration, the list is unsealed first. Then, the new record is added and the list is sealed again. PAL passes the OTP parameters and sealed password list to the Proxy Module which stores sealed password list and sends the OTP parameters to the client. Client generates the same OTP List using the secret phrase and the parameters and outputs the parameters and the OTP list to the user. User can use an OTP application (e.g., mobile phone application) loaded with the parameters provided. Alternatively, he can print the list for manual use.
	
{\bf Proxy Authentication:}	For proxy authentication, the protocol presented in Table \ref{Authentication_Protocol} is executed.


\begin{table}[!htb]
\caption{Authentication Protocol}
\label{Authentication_Protocol}
\centering
\renewcommand{\arraystretch}{1.2}
\begin{tabular}{|l l|}
\hline
1a.Client$\rightarrow$PM &	: Authentication Request\\
2a.PM$\rightarrow$PAL &	: Invoke PAL with "Authentication Block"\\
2b.PM$\rightarrow$PAL &	: sealedPassList\\
3a.Client$\leftrightarrow$PAL &	: Execute Secure Tunnel Protocol\\
3b.Client  &                    : Store PALpub\\
3c.Client$\rightarrow$PAL &	: User ID and Password (master or OTP[n])\\
4a.PAL &		: PassList= TPMUnSeal(sealedPassList,PCR18)\\
4b.PAL &		: validate User ID and Password from PassList\\
4c.PAL &		: If (valid) TPMExtend("1",PCR18) else TPMExtend("0",PCR18)\\
5a.PAL$\rightarrow$PM &	: Validation Result (fail or success)\\
6a.PM	&	: Execute Attestation Protocol\\
6b.PM$\rightarrow$Client &	: If validation = "fail", send error and abort\\
\hline

\end{tabular}
\end{table}

Receiving an authentication request from client, Proxy Module invokes a PAL session, initiates Authentication Block and provides the sealed password list as input. Before password is sent, a secure tunnel between client and the PAL is established. Client stores the received PAL public key used in secure tunnel establishment for later use.

User enters his user ID and password which is then sent to the PAL through the secure tunnel.  PAL performs an unseal operation and validates the password. It extends PCR18 with "1" indicating the success of the operation, otherwise it extends PCR18 with "0". It also writes the validation result (fail or success) into the output file. Receiving this output, Proxy Module executes Attestation Protocol and checks the value of PCR18 for validation. If the validation result is fail, it sends an error message to the client and aborts. Otherwise, authentication is successfully achieved.

{\bf Credential Enrollment:} For credential enrollment, the protocol presented in Table \ref{Credential_Enrollment_Protocol} is executed.


\begin{table}[!htb]
\caption{Credential Enrollment Protocol}
\label{Credential_Enrollment_Protocol}
\centering
\renewcommand{\arraystretch}{1.2}
\begin{tabular}{|l l|}
\hline
1a.User$\rightarrow$PM &  	: SSL web site connection request\\
2a.PM & 		: Check whether user is authenticated\\
3a.PM$\leftrightarrow$User & 	: set up SSL connection with the User\\
4a.PM$\leftrightarrow$Server & 	: validate Target Certificate and set up SSL connection with the Target Server\\
5a.PM &		: is login page?\\
6a.PM$\leftrightarrow$DB &	: does user have enrolled credentials?\\
7a.PM$\rightarrow$User &	: If not, send login page with empty fields\\
8a.User &	                  : enter credentials\\
9a.Browser &	: encCredwithPal= Enc(credentials)PALpub\\
10a.Browser$\rightarrow$PM &  	: encCredwithPal\\
11a.PM$\leftrightarrow$PAL &        : Execute Credential Decryption Protocol\\
12a.PM	&	: Insert original credentials\\
13a.PM$\rightarrow$Server &	: Submit credentials\\
14a.Server$\rightarrow$PM &	: User is authenticated\\
15a.PM$\rightarrow$DB &	: store encCredwithPal, sealedPALpriv and nonce\\
16a.PM$\rightarrow$User &	: User is authenticated\\
\hline

\end{tabular}
\end{table}

As we mentioned previously, Proxy Module runs as an SSL MITM and intercepts the SSL connection of the user. First, PM checks whether the user is authenticated. Then, it sets up SSL connections, one for client and one for the target server. Before the SSL connection with the target server, PM validates the target server's certificate. If the visited web page is a login page, PM runs a query in User DB to understand whether the user has previously enrolled credentials for the target web site. If not, it forwards the login page to the user with empty credential fields. Receiving this login page the user enters credentials and clicks the submit button. The browser add-on of client software recognizes that user has filled credential fields and encrypts them with PAL public key which has been stored during the proxy authentication.

Receiving the encrypted credentials, PM first executes Credential Decryption Protocol in order to obtain the plaintext credentials. Then, it inserts the credentials into the relevant fields on the login page and performs submission. If the user is successfully authenticated to the server, PM completes the enrollment protocol by storing the encrypted credentials, sealed PAL private key, nonce and the other user information into the user database.

{\bf Credential Submission:} Credential submission protocol is given in Table \ref{Credential_Submission_Protocol}.


\begin{table}[!htb]
\caption{Credential Submission Protocol}
\label{Credential_Submission_Protocol}
\centering
\renewcommand{\arraystretch}{1.2}
\begin{tabular}{|l l|}
\hline
1a.User$\rightarrow$PM &  	: SSL web site connection request\\
2a.PM &  		: Check whether user is authenticated\\
3a.PM$\leftrightarrow$User & 	: set up SSL connection with the User\\
4a.PM$\leftrightarrow$Server &  : validate Target Certificate and set up SSL connection with the Target Server\\
5a.PM &		: is login page?\\
6a.PM$\leftrightarrow$DB &	: does user have enrolled credentials?\\
7a.PM &		: If yes, create dummy credentials (DumCred)\\
8a.PM$\rightarrow$User & 	: insert DumCred and send login page\\
9a.User$\rightarrow$PM &	: submit login page\\
10a.PM$\leftrightarrow$DB &	: Retrieve user's encCredwithPAL, nonce and sealedPALpriv\\
11a.PM	&                  : Execute Credential Decryption Protocol\\
11b.PM	&	: Insert original credentials\\
12a.PM$\rightarrow$Server & : submit credentials\\
13a.Server$\rightarrow$PM &	: User is authenticated\\
14a.PM$\rightarrow$User	& : User is authenticated\\
\hline

\end{tabular}
\end{table}

Credential Submission Protocol starts as same as the Credential Enrollment Protocol. But, if the credential of the authenticated user for the target web site has already been enrolled in user DB, PM generates dummy credentials and inserts them into the credential fields of login page and sends it to the user. We note that the original credentials are not sent for security reasons. Upon user's click on the Submit button, PM retrieves the encrypted credentials of the user, sealed PAL private key and nonce values from user database, executes Credential Decryption Protocol and obtains the plaintext credentials. Then, it inserts the credentials and completes the submission operation by submitting the login page to the target server.

{\bf Credential Update:} The protocol for credential update is given in Table \ref{Credential_Update_Protocol}.


\begin{table}[!htb]
\caption{Credential Update Protocol}
\label{Credential_Update_Protocol}
\centering
\renewcommand{\arraystretch}{1.2}
\begin{tabular}{|l l|}
\hline
1a.User$\rightarrow$PM &  	: SSL web site connection request\\
2a.PM & 		: Check whether user is authenticated\\
3a.PM$\leftrightarrow$User & 	: set up SSL connection with the User\\
4a.PM$\leftrightarrow$Server & 	: validate Target Certificate and set up SSL connection with the Target Server\\
5a.PM	&	: is update page?\\
6a.PM$\leftrightarrow$DB &	: does user have enrolled credentials?\\
7a.PM	&	: If yes, create dummy credentials (DumCred)\\
8a.PM$\rightarrow$User & 	: insert DumCred in old credentials field and send update page\\
9a.User &	                  : enter new credentials (NewCred)\\
10a.Browser &               : encNewCredwithPal=Enc(NewCred)PALpub\\
11a.Browser$\rightarrow$PM &  	: encNewCredwithPal\\
12a.PM &	                  : Execute Credential Decryption Protocol and get NewCred\\
13a.PM$\leftrightarrow$DB &	: Retrieve user's encCredwithPAL, nonce and sealedPALpriv\\
14a.PM &		: Execute Credential Decryption Protocol and get Old Credentials (OldCred)\\
14b.PM &		: Insert OldCred and NewCred\\
15a.PM$\rightarrow$Server &	: Submit Update Page\\
16a.Server$\rightarrow$PM & : User credentials are updated\\
17a.PM$\rightarrow$DB &	: Update user's record with the new encNewCredwithPal, sealedPALpriv and nonce\\
18a.PM$\rightarrow$User & 	: User credentials are updated\\
\hline

\end{tabular}
\end{table}

Credential Update Protocol is similar to Credential Submission Protocol. If the visited web site is a password update page, PM checks the User DB to understand whether the user has previously enrolled credentials for the target site. If so, it sends the update page to the user by filling in the old credential fields with the dummy credentials. Then, user fills in only the new credentials part on the update page. The new credentials are encrypted with the PAL public key. Receiving the encrypted credentials, PM first executes Credential Decryption Protocol and obtains the new credentials in plaintext. Then, it retrieves the encrypted old credentials and executes Credential Decryption Protocol to obtain the plaintext old credentials of the user. PM inserts the old and new credentials into the relevant fields on update page and submits the page to the target server. If the update operation on server side is successfully completed, PM updates the user database with the new credentials and new sealed PAL private key and the nonce value.

\subsection{Implementation Details}

In this section, we give brief information on our prototype implementation of Trust-in-the-Middle system.

In our prototype system, we modify and use an SSL MITM Proxy software publicly available \cite{19}. We use OpenSuse 11.2 operating system on an HP DC7800 having a TPM v1.2 chip and Intel TXT support \cite{16}.  For TPM DRTM operations, we use Flicker v.0.2 \cite{9}. For trusted boot we use tboot \cite{18}. For the client software which runs registration, attestation and tunnel (SOCKS) establishment, we use a modified Putty SSH client software \cite{20}. The reason of using a second tunnel besides SSL tunnels in our proposed solution is the need to set up browser independent tunnels where all the web traffic is forwarded through. In order to open an SSH SOCKS tunnel, we configure specific settings on Putty software \cite{21}.

\begin{figure}[!htb]
\centering
\includegraphics[width=3.5in]{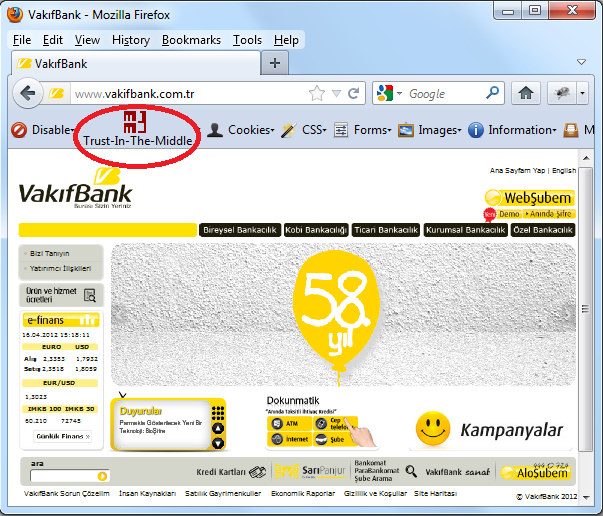}
\caption{Trust-in-the-Middle Browser Add-on}
\label{BrowserAddon}
\end{figure}

For one time passwords, we use OPIE (One-time Passwords In Everything) \cite{22} which is an S/Key One-Time Password implementation.  On client side, we use 1Key which is an OPIE iPhone application \cite{23} for one-time passwords. As the browser add-on, we modify and use a Firefox add-on, Proxy Switch \cite{24}. With this add-on (see Figure \ref{BrowserAddon}) users could configure the proxy settings automatically. This add-on is also responsible for performing encryption operations using the PAL public key.

\section{SECURITY ANALYSIS}
In this section, security analysis of the proposed system is presented. As mentioned, the protection of user credentials on the proxy is our main focus. However, we also mention some of the client and network side issues as well.

\subsection{Security of Tunnel with PAL}
Before a sensitive data is sent to PAL, a Secure Tunnel protocol is executed. This protocol uses TPM DRTM functionality and the attestation protocol. During the PAL session, the generated output including the public key of PAL is extended into PCR18 and the private key of PAL is sealed. By verifying the PCR 18 value, the attestation operation ensures that the correct PAL has been executed and the received public key belongs to this PAL. After this verification, sensitive data is sent encrypted by PAL public key. This encrypted data can only be decrypted by PAL private key in TPM DRTM protection. Since PAL private key is protected by TPM seal operation, it is ensured that only the same PAL, running in TPM DRTM protection, can unseal the private key. As a result, it is ensured that sensitive data cannot be accessed either on network or in proxy once it has been encrypted on client side.

\subsection{Security of the Services}

Preventing disclosure of user credentials to a malicious code running on the proxy system is mandatory to establish the trustworthiness of the system. Credentials are under threat during (i) registration (ii) authentication (iii) enrollment, (iv) update and (v) submission phases:

{\bf Registration.} During registration, master password and the secret phrase used in OTP generation are sent to PAL through the secure tunnel. OTP generation is performed by PAL in TPM DRTM protection and the passwords are stored after being sealed in PAL session.  Only in TPM DRTM protection and in a correct PAL session, passwords can be unsealed.

{\bf Authentication.} Password for proxy authentication is sent to PAL through the secure tunnel. The authentication is performed in PAL and in TPM DRTM protection. Therefore, authentication process cannot be intervened by any other entity. The authentication result is not only written in output file but also extended into PCR18. Hence, proxy module can verify the authentication result by performing attestation.

{\bf Enrollment.} The credentials are encrypted with PAL public key. Since the public key has been verified we ensure that the encrypted credentials can only be accessed on the proxy by PAL. PM runs Credential Decryption protocol which outputs the credentials by encrypting them with PM's public key obtained after a TPM Unseal operation. So this public key is verified  to be the one created during trusted boot with the Initial Sealing protocol. A malicious software cannot obtain the plaintext credentials because they are encrypted with the public key of the proxy module.

{\bf Submission.} Submission operation is performed only if the user is authenticated to the proxy and the target certificate is verified. Encrypted credentials can only be decrypted with the unsealed PAL private key which is available to the correct PAL in TPM DRTM environment. User credentials are sent to proxy module by encrypting them with the public key of proxy which has been created during trusted boot and protected by initial sealing operation.

{\bf Update.} Security of credential update is achieved similarly as in submission and enrollment operations.

\subsection{Modification of Proxy Module}
Modification of Proxy Module is a serious threat. We give below different attack scenarios and analyze how Trust-in-the-Middle provides protection.

\emph{Malicious code infects proxy module just before system reboot.} If proxy module has been maliciously modified before system reboot, trusted boot can detect that the hash of the proxy module has changed and aborts the booting.

\emph{Malicious code stops proxy service and infects proxy module.} If proxy module has been modified after system boot, the final hash value of PCR18 would be different after the integrity measurements and Extend operations in PAL execution. As a result, the attestation fails and sensitive data of the user is not conveyed to the proxy during registration or authentication. Without the authentication, the tunnel cannot be established and the user cannot proceed using the Trust-in-the-Middle system.

\emph{Malicious code stops proxy service and runs a malicious copy of the proxy module.} In this scenario, we assume that malicious code does not change the original code of proxy therefore the attestation may be successful during registration or authentication. However, malicious code cannot access user credentials in storage as they are encrypted with PAL public key attested by the client. It cannot access the plaintext credentials during credential enrollment, submission or update because the credentials conveyed between client and proxy are encrypted with PAL public key. After credentials are decrypted in PAL session, they are encrypted with the public key of the genuine proxy which was created and sealed during trusted boot. So malicious proxy module again cannot have access to the plaintext credentials.

\emph{Malicious code trying to modify sealed public key of proxy module.} Trusted boot ensures the integrity of proxy module. When proxy module has been started, it executes initial sealing protocol. During this protocol, PM generates an RSA key pair, extends the public portion into PCR15.  It then sends the public portion to PAL to seal it. Since the PCR15 value is firstly extended by the genuine proxy module during trusted boot, any malicious module extending the value of PCR15 cannot make PAL to use its public key for sealing because PCR15 verification would fail (since PCR values cannot be set to a specific value, they can only be extended).

\subsection{Modification of the Flicker Module}
Boot time modification of Flicker module can be detected by trusted boot. Load time modification of Flicker Module breaks the trust chain and leads to a failure in unseal operation. It can also be detected by the attestation.

\subsection{Modification of PAL}
If Flicker module loads a modified PAL, PCR18 would have a different hash value after TPM DRTM operation, which results in failure in seal/unseal operations and  attestation.

\subsection{Modification of PCRs}
TPM DRTM ensures that no other operation can reset the value of PCRs to zero and TPM guarantees that PCR values cannot be set to a default value and can be written only by the TPM Extend operation.

\subsection{Modification of Input}
The modification of input leads to a denial of service attack but does not reveal credentials.  If the input value is wrong, seal/unseal operations would fail. PAL cannot recover the required data and abort.

\subsection{Modification of Output}
Critical output values are extended into PCR18 by PAL and sent in the SML. Hence, the output values are verified in attestation.

\subsection{Replay Attacks}
For each authentication session a new RSA key pair is created by PAL and a nonce value is used in private key seal operation.

A nonce is used in encrypting the credentials with proxy public key in credential decryption protocol. So the encrypted passwords cannot be replayed.

\subsection{Run-time and Memory Based Attacks}
TPM DRTM environment guarantees that PAL is executed in an isolated environment which cannot be intervened by any malicious entity. When PAL quits, Intel TXT ensures that all relevant memory locations are cleaned before exiting. However, run-time memory based attacks to the proxy module capturing its private key or user credentials are possible. In order to prevent this attack, the credentials can be encrypted in the PAL session with public key of the target server. However, this protection violates the transparency requirements as it requires change on server side. Run-time software integrity problems were studied in detail and several solutions were available in the literature \cite{25,26,27,28,29,30}. We consider to incorporate these solutions into the Trust-in-the-Middle system in our future work.
	
\subsection{Client and Network Side Threats}
The proposed system supports one-time passwords to provide protection against keyloggers, screenscrapers and any other malware on the client environment trying to capture user credentials. However,  the user is still under threat from more sophisticated malware such as transaction generators \cite{31}.

Since Trust-in-the-Middle encrypts all the sensitive data transmitted in the network, network attacks aiming to capture the plaintext credentials are prevented.

With the assumption that certificates of the target servers are retrieved securely by the proxy and it is checked before setting up SSL connections,  server-forging attacks could not be conducted.

\section{CONCLUSION and FUTURE WORK}
In this paper, we present Trust-in-the-Middle, a TPM based proxy system for secure and usable management of authentication credentials. Our goal is to increase the security of the proxy system in order to make it a trustworthy central intermediary system which takes over credential storage and submission operations from its users.

Using TPM DRTM functionality, Trust-in-the-Middle securely authenticates users and stores their credentials on proxy encrypted with TPM protected keys. Whenever these credentials need to be used, they are securely decrypted and submitted to the target servers.

Security critical operations on the proxy are performed in an isolated environment protected by TPM DRTM and credentials are never disclosed to outside without being encrypted. Attestation is used to verify the integrity of security sensitive code running in TPM DRTM protection and the modules running on proxy. Only if the verification succeeds, then the sensitive data is sent to these modules.	In this work, we report on a still progressing work for increasing the security of authentication proxies using Trusted Computing technologies. Below, we provide future work directions together with the technical limitations in the current proposal.

In the current version of Trust-in-the-Middle, run-time integrity problems are not taken into consideration. This is a serious limitation considering the security of user credentials on proxy. Therefore, adopting run-time security measures in our proposed system is the first item in our research agenda.

Our proof-of-concept implementation requires a special software installation on user side which may be considered as not being a usable solution. We think that not only this single aspect but the whole user experience of the system should be rigorously evaluated with carefully-crafted usability studies.

Proxy systems are used not only for user authentication but also for many other security and privacy purposes (e.g., \cite{32}). Since proxy trust problem is common in all of these applications, we think it is a promising future work to investigate on positioning our solution as a more general framework.

In closing, we recognize trust concerns as a major obstacle to voluntary widespread use of proxy systems. It is a major challenge to build systems which are secure, trusted and usable.  Our proposed system, Trust-in-the-Middle, is a step toward meeting this challenge.

%

\bibliographystyle{IEEEtran}
\bibliography{IEEEabrv,Trust-in-the-Middle}

\end{document}